\newcommand{\ale}{\ \raisebox{-.3ex}{$\stackrel{<}{\scriptstyle \sim}$}\ }

\documentclass{mn2e}

\input psfig.sty

\title[Protostellar disc fragmentation]
	{Substellar companions and isolated planetary mass objects from protostellar disc fragmentation}

\author[Rice et al.]{W.K.M. Rice$^1$, P.J. Armitage$^{2,3}$, I.A. Bonnell$^1$, M.R. Bate$^4$, S.V. Jeffers$^1$, S.G. Vine$^1$ \\
	$^1$School of Physics and Astronomy, University 
	of St Andrews, North Haugh, St Andrews KY16 9SS \\
	$^2$JILA, Campus Box 440, University of Colorado, Boulder CO 80309-0440, USA \\
	$^3$Department of Astrophysical and Planetary Sciences, University of Colorado, Boulder CO 80309-0391, USA \\ 
	$^4$School of Physics, University of Exeter, Stocker Road, Exeter EX4 4QL}		
\begin{document}

\maketitle

\begin{abstract} 
Self-gravitating protostellar discs are unstable to fragmentation if the 
gas can cool on a time scale that is short compared to the orbital period. 
We use a combination of hydrodynamic simulations and N-body orbit integrations 
to study the long term evolution of a fragmenting disc with an initial mass 
ratio to the star of $M_{\rm disc} / M_* = 0.1$. For a disc which is initially 
unstable across a range of radii, a combination of collapse and 
subsequent accretion yields substellar objects with a spectrum of masses 
extending (for a Solar mass star) up to $\approx 0.01 \ M_\odot$. 
Subsequent gravitational evolution ejects most of the lower mass 
objects within a few million years, leaving a small number of very 
massive planets or brown dwarfs in eccentric orbits at moderately 
small radii. Based on these results, systems 
such as HD~168443 -- in which the companions are close to or beyond the 
deuterium burning limit -- appear to be the best candidates to have formed 
via gravitational instability. If massive substellar companions originate from
disc fragmentation, while lower-mass planetary companions originate from core
accretion, the metallicity distribution 
of stars which host massive substellar companions at radii of 
$\sim 1 {\rm au}$ should differ from that of stars with lower 
mass planetary companions.
\end{abstract}


\section{Introduction}
Protostellar discs formed during the early phases of star formation 
can be cool and massive enough that self-gravity plays an 
important role in their evolution (Cassen \& Moosman 1981; 
Lin \& Pringle 1990; Bate, Bonnell \& Bromm 2003; though see 
also Krasnopolsky \& K\"onigl 2002). Self-gravity leads to the 
formation of spiral structure, which in turn can drive angular 
momentum transport and accretion (Adams, Ruden \& Shu 1989; 
Laughlin \& Bodenheimer 1994; Laughlin \& 
Rozyczka 1996; Laughlin, Korchagin \& Adams 1997; 
Pickett et al. 1998; Nelson et al. 1998; Nelson, Benz \& Ruzmaikina 2000).
Alternatively, a sufficiently unstable disc may fragment into bound 
objects, which in a protostellar disc would have planetary or 
brown dwarf masses.

Whether a gravitationally unstable gas disc will fragment or stably transport 
angular momentum depends upon the efficiency of radiative cooling from 
the disc's surfaces (Pickett et al. 2000, 2003; Gammie 2001; Boss 2001, 2002b; 
Rice et al. 2003). Efficient disc cooling -- on a time scale 
comparable to the orbital period -- robs transient overdensities of 
pressure support and allows them to collapse into bound substellar 
objects. Quantitatively, Gammie (2001) showed using analytic considerations 
and local numerical simulations that fragmentation occurs if  
the local cooling time is $\ale 3 \Omega_K^{-1}$, where $\Omega_K$ is 
the Keplerian angular velocity. A similar fragmentation boundary was 
obtained in global simulations of cooling, self-gravitating discs (Rice et al. 2003).
For discs whose mass is substantially smaller than that of the star, 
fragmentation leads immediately to the formation of a number of 
substellar objects (Mayer et al. 2002; Rice et al. 2003).

The conditions in the outer regions of protostellar discs at early 
epochs are not well known, so it is uncertain whether the rapid 
cooling required for fragmentation occurs frequently, rarely, or 
never. Here, we assume that the conditions in the disc are such that 
fragmentation occurs, and investigate the subsequent evolution 
of the system, which will initially comprise a number of substellar 
objects embedded within the remaining disc gas. Qualitatively, it 
is fairly clear how such a system evolves post-fragmentation. The gas 
will be accreted -- either by the planets or by the star -- while 
gravitational interactions amongst the planets or brown dwarfs  
will eject most of them while leaving a handful of survivors 
on eccentric orbits (e.g. Armitage \& Hansen 1999). Numerical 
integrations of unstable multiple planet systems (Lin \& Ida 1997; 
Papaloizou \& Terquem 2001; Terquem \& Papaloizou 2002, Adams \& Laughlin 2003) show 
that the final planetary systems can be strikingly similar to 
some of those observed in radial velocity surveys (Marcy \& 
Butler 1998), though extremely close in planets such as that 
orbiting 51~Peg (Mayor \& Queloz 1995) almost certainly require 
additional migration mechanisms (Lin, Bodenheimer \& Richardson 1996).

In this {\em letter}, we study the long term evolution of a 
fragmenting protostellar disc using a two step approach. 
We first extend our previous hydrodynamic simulation of 
a fragmenting disc (Rice et al. 2003) until most of the 
gas has been swept up and incorporated into bound 
objects. This allows us to estimate the mass spectrum 
of substellar objects produced as a consequence of 
disc fragmentation. We then isolate the population of 
substellar objects, and evolve them under purely 
gravitational forces (using methods similar to Papaloizou \& 
Terquem 2001, and obtaining comparable results) until a 
stable final system is obtained. In doing this we are able
to self-consistently study the formation and evolution of a
planetary system. 

\section{Methods}
We use smoothed particle hydrodynamics (SPH) (e.g., Benz 1990; Monaghan 1992) to consider 
a $0.1 M_{\rm \odot}$ disc, with a radius of $50$ au, 
surrounding a $1 M_{\rm \odot}$ star. The disc is modelled, in three-dimensions, 
using 250000 SPH particles, while the star
is represented by a sink particle onto which gas particles may accrete if they approach to
within an accretion radius of $0.5$ au (Bate, Bonnell \& Price 1995). 
Disc self-gravity is included and a tree is used to determine gravitational
forces between gas particles and between gas particles and point masses. The 
gravitational force between point masses is computed directly.

The disc temperature ($T$) and surface density ($\Sigma$) are taken, 
initially, to have radial profiles of $T \propto r^{-0.5}$ and $\Sigma \propto r^{-1}$. 
The temperature is normalised to give a minimum Toomre (1964) $Q$ parameter of $2$ at the outer disc edge.
Since the disc stability depends on both heating
and cooling we use an adiabatic equation of state, with adiabatic index $\gamma=5/3$, and impose a 
radially dependent cooling time. The imposed cooling time has the form 
$t_{\rm cool} = 3 \Omega^{-1}$, where $\Omega$ is the angular frequency. The motivation for choosing this
cooling time is provided by
both local and global simulations (Gammie 2001; Rice et al. 2003) which show that a self-gravitating accretion disc
will fragment into gravitationally bound objects for $t_{\rm cool} \le 3 \Omega^{-1}$.
The form of this imposed cooling time can be related, at least approximately, to the real physics of an accretion
disc. For an optically thick accretion disc in equilibrium it can be shown 
(e.g., Pringle 1981) that the cooling time is given by
\begin{equation}
t_{\rm cool} = \frac{4}{9 \gamma (\gamma - 1)} \frac{1}{\alpha \Omega}
\end{equation}
where $\gamma$ is the adiabatic index, and $\alpha$ is the Shakura \& Sunyaev (1973) viscosity parameter. 

The fragmentation of the disc produces gravitationally bound regions with densities significantly higher
than the initial disc density. Continuing to follow the internal evolution of these fragments -- which
cannot be done reliably in any case -- tends to slow the code down significantly. 
To continue simulating
the fragmenting disc we allow sufficiently dense regions, containing $\sim 50$ SPH particles,  
that are gravitationally bound to be converted into sink particles (Bate, Bonnell \& Price 1995). Since we are
simulating a $0.1 M_{\rm \odot}$ disc using 250000 SPH particles we have a minimum sink particle mass of 
$\sim 2 \times 10^{-5} M_{\rm \odot}$. Gas particles may accrete onto the sink particles if they approach to
within a predefined accretion radius. The accretion radius is taken to be $0.02$ au and is approximately the Hill radius 
of a minimum mass sink particle at $1$ au. We are therefore not only able to simulate the fragmentation of the disc, but
are also able to continue following the subsequent growth of the gravitationally bound fragments. 

\section{Disc evolution}
\begin{figure}
\centerline{\hbox{
\psfig{figure=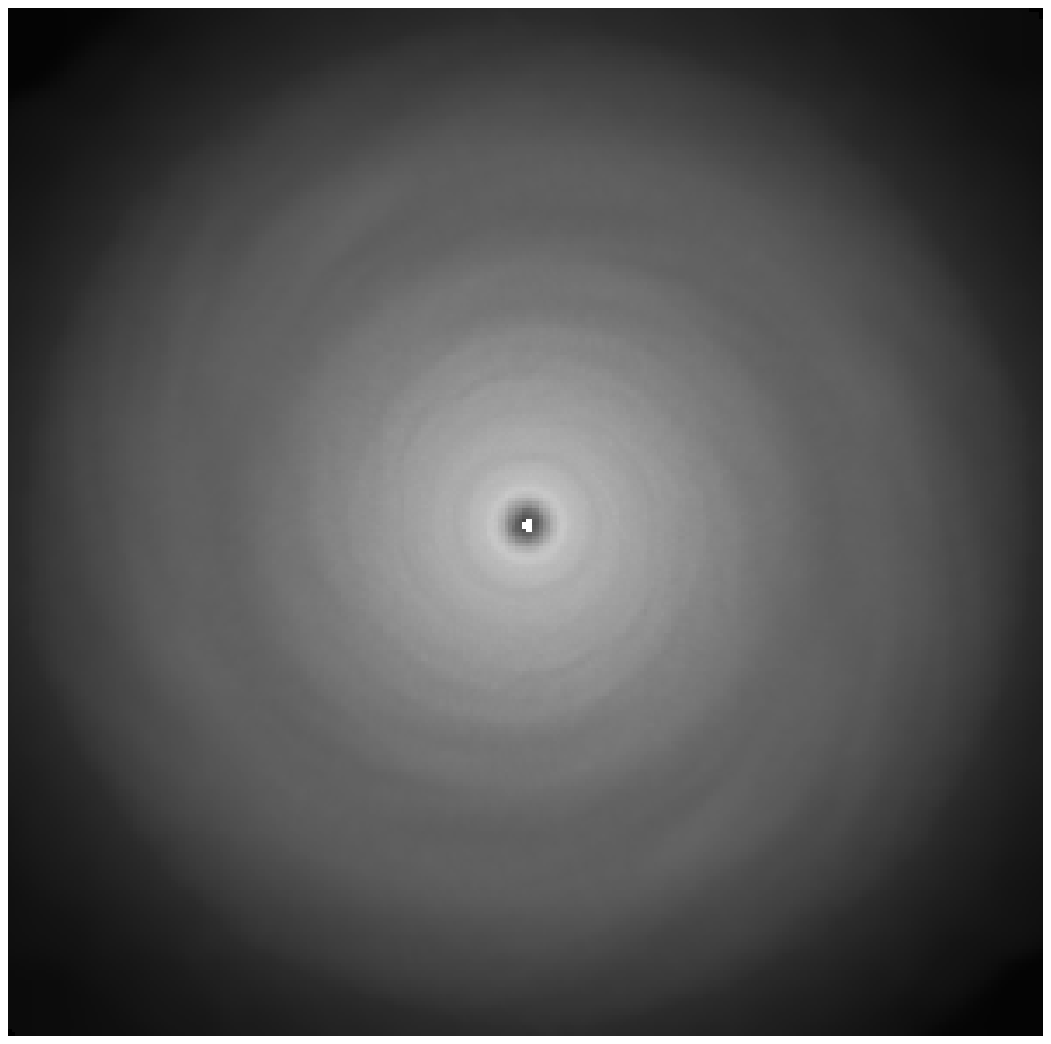,width=1.6truein}
\psfig{figure=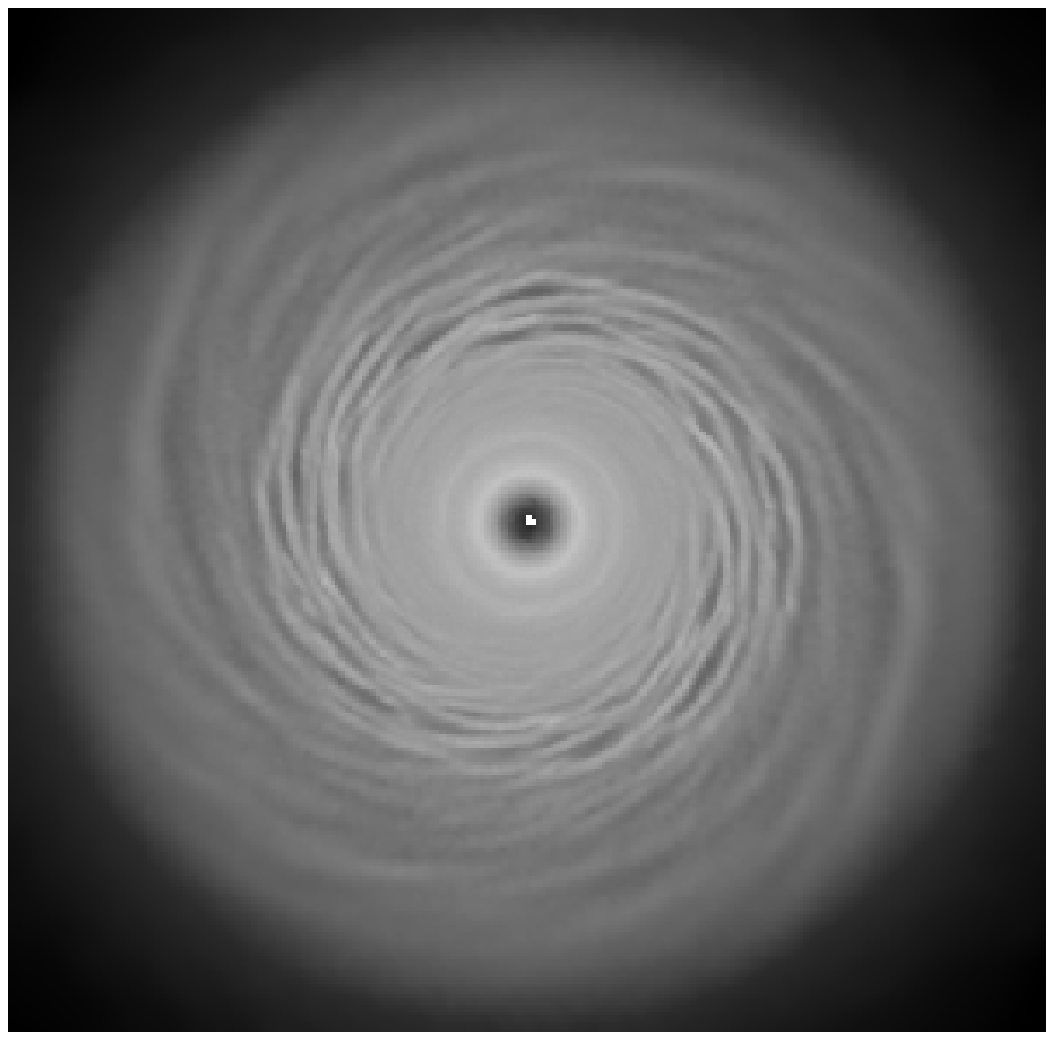,width=1.6truein}}}
\centerline{\hbox{
\psfig{figure=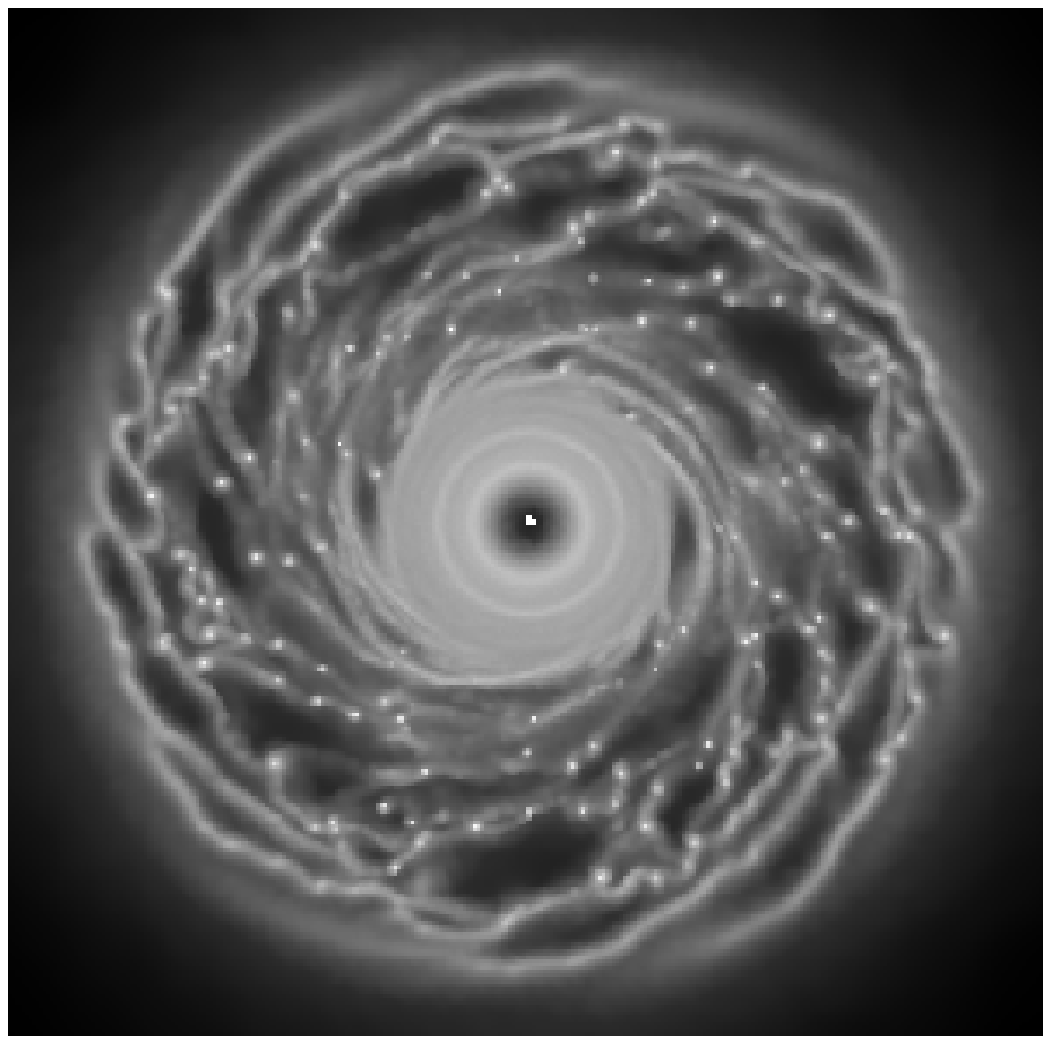,width=1.6truein}
\psfig{figure=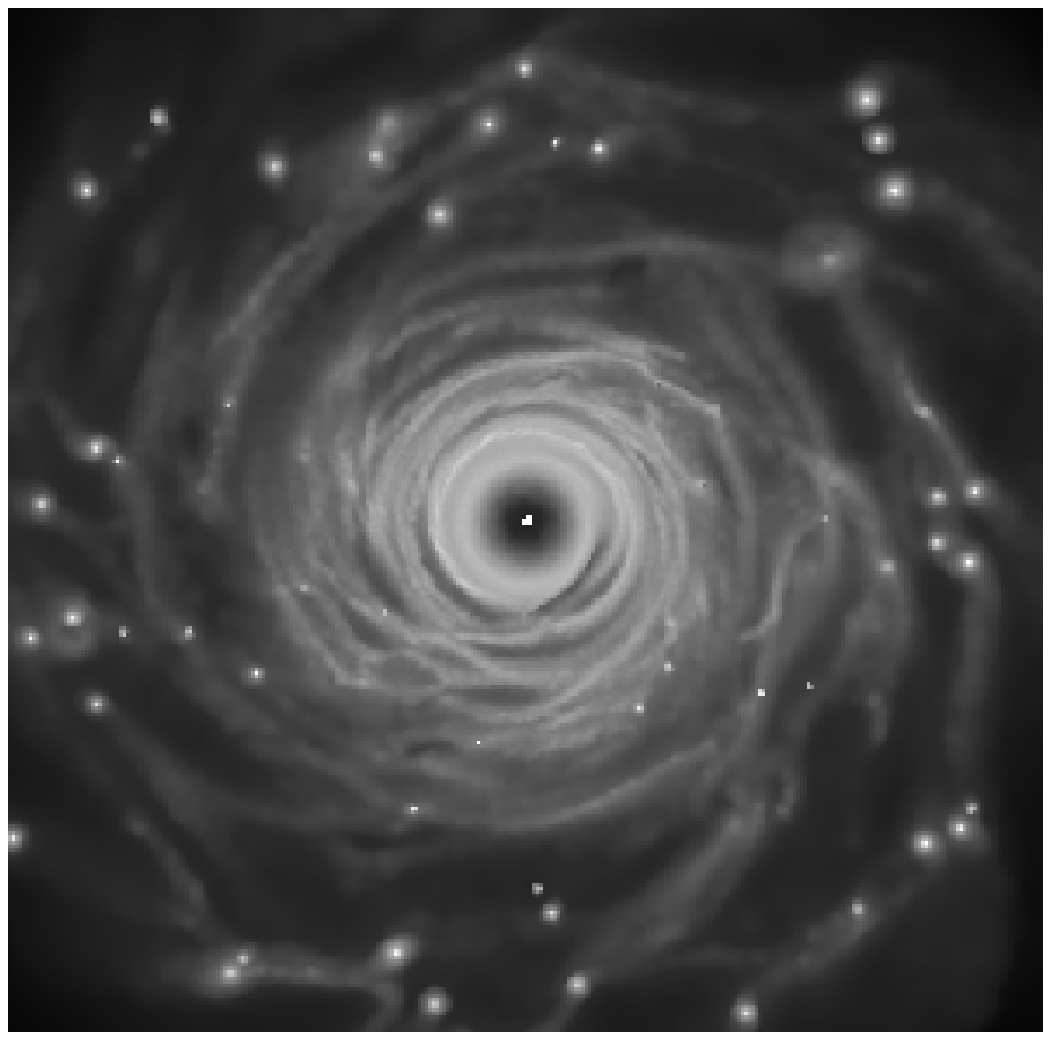,width=1.6truein}}}
\caption{Disc surface densities at four different times during the hydrodynamical simulation. 
After $140$ yrs (top left) the disc is still reasonably smooth and unstructured. As the disc
evolves ($t = 420$ yrs - top right) spiral structures, due to the growth of the gravitational instability,
are evident. The low cooling time means that heating through the growth of the gravitational instability
is unable to balance the imposed cooling without the disc fragmenting. Gravitationally bound fragments are
clearly present after $644$ yrs (bottom left). The bottom right figure shows the disc after $956$ yrs and shows
fragmentation taking place in the outer regions of the disc while the inner fragments have been converted into
point masses.}
\label{hydro}
\end{figure}

Figure \ref{hydro} shows the surface density structure of the gaseous protoplanetary disc at four different times. All four
figures have $x$ and $y$ axes that run from $-60$ au to $60$ au and have the star in the center. At early times ($t = 140$ yrs), 
the surface density is reasonably smooth and unstructured.  As the disc evolves ($t = 420$ yrs) spiral structures, that
are due to the growth of the gravitational instability, are evident. Since the cooling time in this
particular simulation is short, heating through the growth of the gravitational instability is unable
to balance the imposed cooling without the disc fragmenting into gravitationally bound objects. After $644$ years there
is clear evidence of fragmentation with a number of high density regions present in the disc. After $956$ years, there
is still fragmentation of the outer regions of the disc.

We continue to evolve the hydrodynamical simulation for a further $10640$ years, at which point $83$ substellar
objects have formed through fragmentation of the disc gas. All $83$ objects have been converted into
point masses (Bate, Bonnell \& Price 1995), and $87$ $\%$ 
of the gas ($217362$ SPH particles) has been accreted onto either these substellar objects or onto the central star.
The central star has increased in mass from $1 M_{\rm \odot}$ to $1.011 M_{\rm \odot}$, an accretion rate of
$\sim 10^{-6} M_{\rm \odot}$ yr$^{-1}$.
At this stage, we remove the remaining gas and evolve the 84 point masses 
(central star plus 83 substellar objects) using an
N-body code. We primarily use NBODY3 (Aarseth 1999), which is fast and uses chain regularization to
efficiently treat close encounters and binary systems, but also compare our results using hnbody (Rauch \& Hamilton 2002
and Mercury (Chambers 1999).    

An obvious limitation of our simulations is that the derived fragmentation time scale (around $10^4$~yr) 
is significantly shorter than the time scale on which the disc is assembled. Unless instability in 
the disc is radiatively triggered, for example by a sudden change in the illumination of the 
outer disc by the central star, this disparity in time scales implies that disc fragmentation 
should properly be studied within the larger context of the star and disc formation process. 
This is extremely difficult. Although simulations of star formation within molecular 
clouds (Watkins et al. 1998; Bate, Bonnell \& Bromm 2002) already support the view that 
brown dwarfs may form within protostellar discs, they do not yet have the resolution or 
treatment of the thermal physics to follow the fragmentation process in the same detail as 
is possible for an isolated disc.

\section{Substellar mass function}
\begin{figure}
\centerline{\psfig{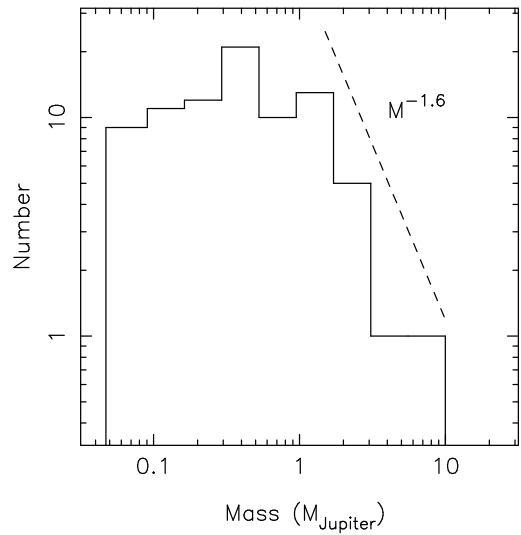}}
\caption{Substellar Initial Mass Function (IMF). There is a turnover at $\sim 10^{-3} M_{\rm \odot}$ with
a slope, above $10^{-3} M_{\rm \odot}$, of dN/dlogM $\propto M^{-1.6}$ and a maximum mass of $7.8 M_{\rm Jupiter}$}
\label{imf}
\end{figure}

Figure \ref{imf} shows the substellar Initial Mass Function (IMF) immediately following disc fragmentation
(i.e., prior to subsequent modification by planet-planet and planet-star collisions). 
Between $\sim 3 \times 10^{-2} M_{\rm Jupiter}$ and 
$\sim 1 M_{\rm Jupiter}$, the IMF is reasonably flat with $\sim 60$ substellar objects having masses less than $1
M_{\rm Jupiter}$. There is a turnover at $\sim 1 M_{\rm Jupiter}$ above which the 
Mass Function falls off steeply with 
dN/dlogM $\propto M^{-1.6}$. In this particular case we have a maximum mass of 
$7.8 M_{\rm Jupiter}$. The slope of the IMF cutoff, and the maximum mass, are likely to depend on the disc properties. Although
we have not performed any kind of parameter survey, a second simulation, with the same total disc mass but a steeper
surface density profile, produced objects that were slightly more massive. 
The maximum mass is also likely to depend on the disc mass.
Our simulation is essentially scale free.  We could, equally well, have assumed a stellar mass 
of $M_* = 2 M_{\rm \odot}$, giving a disc
mass of $0.2 M_{\rm \odot}$ and increasing the masses of the substellar objects by a factor of $2$. The maximum mass would then be
$15.6 M_{\rm Jupiter}$. Similarly, simulations by Boss (1998) show that a more massive disc around a star with the same 
mass ($1 M_{\rm \odot}$) also result in substellar objects with masses in excess of $10 M_{\rm Jupiter}$. 
Since only reasonably massive discs will become sufficiently gravitationally unstable for
fragmentation, it seems likely that the maximum mass of objects produced in discs via fragmentation would also be reasonably high
($\sim 10 M_{\rm Jupiter}$ or greater). 


\section{N-body evolution}
After $11618$ years, $87 \%$ of the disc gas has been accreted onto the $83$ substellar objects, or onto
the central star. At this stage we remove the remaining gas and evolve the $84$ point masses (central star
plus 83 substellar objects) in an N-body fashion using NBODY3 (Aarseth 1999), which we normalise in the
standard way by setting the radius and total mass to $1$. The velocities are then
normalised to recover the original virial ratio. The system is evolved
for $21$ Myrs. Of the $83$ substellar objects, $74$ are ejected from the system and would become, unless captured by
another system, free-floating planets (e.g., Lin \& Ida 1997, Papaloizou \& Terquem 2001, Smith \& Bonnell 2001, 
Terquem \& Papaloizou 2002, Hurley \& Shara 2002). 
Of these $74$ ejected objects, $19$ have masses in excess of $1 M_{\rm Jupiter}$, with
the most massive having a mass of $3.6 M_{\rm Jupiter}$.

\begin{figure}
\centerline{\psfig{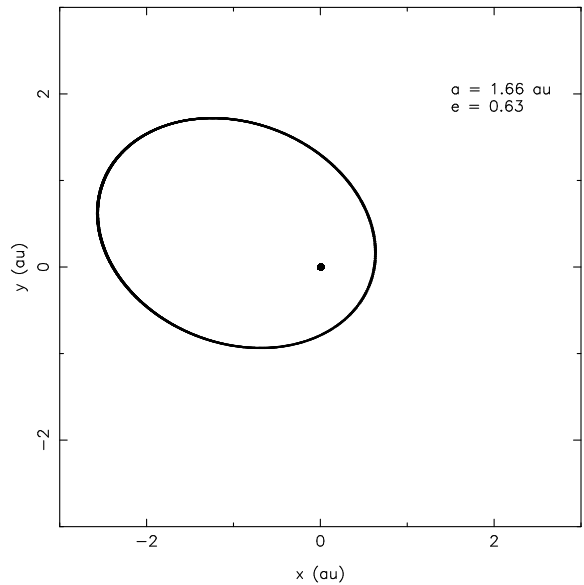}}
\caption{Orbit of the only remaining substellar object within $10$ au. This is the most massive 
object, with a mass of $7.4 M_{\rm Jupiter}$, and has a semi-major axis of $1.66$ au and an eccentricity
of $0.63$.}
\label{orbit}
\end{figure}

Most of the substellar objects that remain bound have large semi-major axes ($a > 500$ au) and 
eccentricities, and (depending on the stellar environment) would probably be 
removed by encounters with other stars. If, however, 
they were able to remain bound, they would be good candidates for direct imaging surveys
for extrasolar planets, either during the late pre-main-sequence phase, around main sequence
stars, or once the central star has evolved into a white dwarf (Burleigh, Clarke \& Hodgkin 2002).
Two of the objects, however, are left on orbits close to the central star. One of these objects
has an orbit that approaches to within a solar radius and so would collide with the star were the stellar
radius included in the calculation. The other, which was the most massive substellar object ($7.8 M_{\rm Jupiter}$), 
had a final semi-major axis of $1.66$ au,
and an eccentricity of $0.63$. These orbital parameters fall well within the range of observed values (Marcy \& Butler 2000). 
Figure \ref{orbit} shows the orbit of this object. It therefore seems that the evolution of a system in which disc fragmentation
produces planetary mass objects, can result in a final state consistent with that currently observed.

The same calculation was also performed using
hnbody (Rauch \& Hamilton 2002) and Mercury (Chambers 1999), in both cases using the Burlisch-Stoer
integrator provided with those codes. Comparable results were obtained from all three codes, 
provided that each was set to model strictly point-mass evolution. In this limit, the most massive
($7.8 M_{\rm Jupiter}$) object was left in an orbit with a semi-major axis of between $1.6$ au and
$1.8$ au, and an eccentricity in excess of $0.3$. We note, however, that this final
result is sensitive to close encounters between planets and between planets and the star. Rerunning
the integration, using Mercury, with reasonable values for the planetary density, and a stellar
radius of $1 R_{\rm \odot}$, we obtained a final orbit for the most massive object that was
significantly wider ($\sim 10$ au). This may be due to the reduced population of low-mass
scatterers, many of which collided with the central star.

As a further test of these N-body results, we performed $10$ N-body simulations in which we 
randomised the positions of the 83 substellar obejcts, keeping the energy and angular 
momentum of each body constant. In $9$ of the $10$ cases, the results were consistent 
with the original N-body calculation. The $7.8 M_{\rm Jupiter}$ body was left in an
orbit with semi-major axis between $1.49$ and $1.81$ au, and eccentricity between $0.28$ and 
$0.82$. The exception was a case in which the $7.8 M_{\rm Jupiter}$ body was ultimately ejected 
from the system while two less
massive objects merged with the central star. In general, however, it does seem that the
final state will be one in which the most massive body remains bound to the star with a 
modest semi-major axis and a reasonably large eccentricity ($\sim 0.3$ or greater).
   
\section{Discussion}
We have used global hydrodynamic simulations to follow the long term fate of a self-gravitating disc that is unstable to
fragmentation (according to the results of Gammie 2001; Rice et al. 2003). For our choice of parameters (disc radius of
50~au, disc mass 0.1~$M_\odot$ around a 1~$M_\odot$ star), we find that fragmentation is largely complete within about
$10^4$ yr. By this epoch, 87~per cent of the disc gas had been accreted, and in excess of 80 substellar objects formed.
The substellar IMF immediately following fragmentation is found to be roughly flat below a Jupiter mass, with a
steep fall off (dN/dlogM $\propto M^{-1.6}$) at higher masses. For our specific simulation, the most massive object had
a mass of 7.8~$M_{\rm Jupiter}$. Since gravitationally unstable discs are likely to have {\em at least} as much mass
as that simulated here, it is likely that the most massive object formed as a result of gravitational
instability will typically be either a very massive planet or a low mass brown dwarf

To continue evolving the system toward an observable epoch, we removed the (small) residual gas fraction and integrated
the multiple planet system using N-body methods. Generically, one expects that the most massive object will often survive
as a bound planet or brown dwarf, while most of the lower mass objects are ejected. This was indeed the outcome.
The 7.8~$M_{\rm Jupiter}$ planet ended up in an orbit with a semi-major axis of 1.5 - 1.8~au and a large
eccentricity; 7 lower mass planets were left as distant compantions ($a > \sim 500$~au); and most
of the rest were ejected. These results are consistent with those of Papaloizou \& Terquem (2001) and
Adams \& Laughlin (2003). They suggest that the extrasolar planetary systems most likely to be the
products of gravitational instability are those in which one or two very massive planets (or low mass
brown dwarfs) orbit at modest radii in highly eccentric orbits. HD~168443 represents an observed example of
such a system (Marcy et al. 2001).

\begin{figure}
\centerline{\psfig{figure=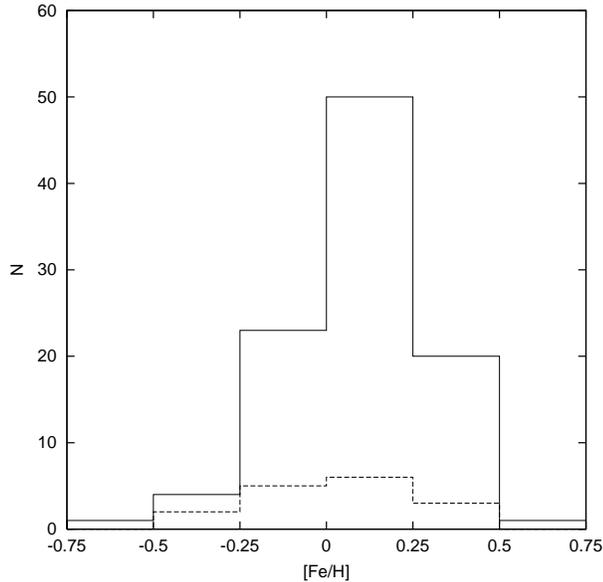,width=3.0truein}}
\caption{Metallicity distribution for all planet-bearing stars (solid line), compared to those that
have at least one companion with a mass in excess of $5 M_{\rm Jupiter}$, a semi-major
axis greater than $0.1$, and an eccentricity greater than $0.2$ (dashed line). It does appear that the
systems preselected on the basis of companion mass, semi-major axis, and eccentricity are not
as metal-rich as planet bearing systems in general.}
\label{metal}
\end{figure}

For observations, our results have three main implications. First, they suggest that gravitational instability, if it
occurs, is likely to populate preferentially the very high mass end of the planetary mass function. There is no reason
to expect that the metallicity dependence of stars hosting massive companions formed via disc fragmentation would be
the same as that for stars with lower-mass planetary companions originating from core accretion (Boss 2002a). 
It is now generally
accepted that stars with planetary companions are generally 
metal-rich (Laughlin 2000, Santos et al. 2001, Murray \& Chaboyer 2002, Santos et al. 2003, Fischer \& Valenti 2003).  
Using the currently available database of known extrasolar planets (see http://cfa-www.harvard.edu/planets/cat1.html), 
we have compared the metallicity distribution for all planet bearing stars with that for systems 
in which there is a companion having a mass in excess of $5 M_{\rm Jupiter}$, a semi-major axis
in excess of $0.1$ au, and an eccentricty greater than $0.2$. The result is shown in Figure \ref{metal} and
illustrates that the systems chosen using the above criteria (dashed line) do not appear to be 
as metal-rich as planet bearing stars in general (solid line). 
Although this is preliminary and inconclusive, a {\em different} metallicity distribution for stars
hosting the most massive planets, compared to stars hosting lower-mass companions, 
would provide circumstantial evidence that massive planets formed from disc
fragmentation. Second, our integrations indicate that additional massive planets could be present at very large orbital
radii in systems with the most massive extrasolar planets or close brown dwarf companions. Although these distant
companions are vulnerable to disruption by encounters with passing stars, they are potentially detectable via
direct imaging. Finally, we find that a significant fraction of the initial gas content of the disc can
end up being ejected from the system in the form of isolated planetary mass objects. Constraining the
numbers -- and especially the masses -- of free-floating substellar objects in star forming regions
is not straightforward for the young objects of relevance here (e.g. Baraffe et al. 2002). 
However, an absence (or small number) of free-floating sub-Jupiter mass planets would provide
a new way to limit the fraction of stars whose discs underwent large scale gravitational collapse of the
sort simulated here.

\section*{Acknowlegments}

The simulations reported in this paper made use of the 
UK Astrophysical Fluids Facility (UKAFF). WKMR and SGV acknowledge support from
PPARC standard grants. SVJ acknowleges support from PPARC and from a University of 
St Andrews scholarship. This paper is based in part upon work
supported by NASA under Grant NAG5-13207 issued through the Office of
Space Science. The authors would like to thank Gregory Laughlin for a useful suggestion,
John Chambers for the use of {\sc mercury} and
for useful discussions, David Asher for help with using {\sc mercury},
and Kevin Rauch and Doug Hamilton for the use of their hnbody integrator.

\end{document}